\documentclass[journal]{IEEEtran}
\usepackage{amsmath,amssymb,amsthm,bm}
\usepackage{graphicx}
\usepackage{booktabs}
\usepackage[hidelinks]{hyperref}
\usepackage{cite}
\usepackage{algorithm}
\usepackage{orcidlink}
\usepackage{tikz}
\usepackage{algpseudocode}
\makeatletter
\renewcommand\fs@ruled{\def\@fs@cfont{\rmfamily}\let\@fs@capt\floatc@ruled
  \def\@fs@pre{\hrule height.8pt depth0pt \kern2pt}%
  \def\@fs@post{\kern2pt\hrule\relax}%
  \def\@fs@mid{\kern2pt\hrule\kern2pt}%
  \let\@fs@iftopcapt\iftrue}
\makeatother
\algrenewcommand\algorithmicrequire{\textit{Input:}}
\algrenewcommand\algorithmicensure{\textit{Output:}}
\algrenewcommand\algorithmicrepeat{\textit{repeat}}
\algrenewcommand\algorithmicuntil{\textit{until}}
\algrenewcommand\algorithmicfor{\textit{for}}
\algrenewcommand\algorithmicdo{\textit{do}}
\algrenewcommand\algorithmicend{\textit{end}}
\newtheorem{proposition}{Proposition}
\newcommand{\bo}{\mathbf{o}}
\newcommand{\bmu}{\bm{\mu}}
\newcommand{\bSig}{\bm{\Sigma}}
\newcommand{\Nc}{\mathcal{N}}
\newcommand{\dTV}{d_{\mathrm{TV}}}
\newcommand{\E}{\mathbb{E}}

\definecolor{lime}{HTML}{A6CE39}
\DeclareRobustCommand{\orcidicon}{%
	\begin{tikzpicture}
	\draw[lime, fill=lime] (0,0)
	circle [radius=0.16]
	node[white] {{\fontfamily{qag}\selectfont \tiny ID}};
	\draw[white, fill=white] (-0.0625,0.095)
	circle [radius=0.007];
	\end{tikzpicture}
	\hspace{-2mm}
}

\foreach \x in {A, ..., Z}{%
	\expandafter\xdef\csname orcid\x\endcsname{\noexpand\href{https://orcid.org/\csname orcidauthor\x\endcsname}{\noexpand\orcidicon}}
}

\begin{document}

\title{Cross-Subject Priors for Maximum a Posteriori Estimation of Latent Structure Influence Models in Motor-Imagery EEG With Few Trials}

\author{Sagnik~De~\orcidA{} and Anil~Kumar~Gupta~\orcidB{}
\thanks{Manuscript received Aug XX, 20XX;
revised XXXX, 20XX; accepted XXXX, 20XX.
Date of publication XXX XX, 20XX; date of current version XXX XX, 20XX. 
\textit{(Corresponding author: Sagnik De)}}
\thanks{S.~De is with the Institute of Radio Physics \& Electronics, University of Calcutta, Kolkata 700009, India.}
\thanks{A.~K.~Gupta is with the HPC Infrastructure and EcoSystem Group, Centre for Development of Advanced Computing, Pune 411008, India.}
\thanks{Digital Object Identifier  10.1109/XXXX.202X.XXXXXXX}}

\maketitle

\pagestyle{headings}

\markboth
{}
{Your Odd Page Header}

\begin{abstract}
    
Coupled hidden Markov models such as the latent
structure influence model (LSIM) describe how the sensorimotor rhythms of the two
hemispheres evolve and interact during motor imagery, but subject-specific maximum
likelihood (ML) estimates are unstable when few trials are available. This paper derives a
maximum a posteriori (MAP) estimator for the LSIM whose conjugate prior equals the expected
statistics of $\bm{\lambda}$ pseudo-trials from a population model fitted to other subjects.
The M-step remains in closed form, expectation maximization keeps its monotone ascent, and
ML ($\bm{\lambda=0}$) and the population model ($\bm{\lambda\to\infty}$) are limiting cases.
The prior strength was selected on 35 subjects of the PhysioNet EEG Motor Movement/Imagery
Dataset and applied unchanged to 70 held-out subjects and to BCI Competition IV-2a. With
$\bm{\lambda=1}$ the held-out log-likelihood exceeded that of ML in 61 of 70 subjects
(Holm-adjusted $\bm{p=9.1\times10^{-8}}$) and in all 70 subjects with five training trials
per class. A prior learned only from the PhysioNet data improved the held-out likelihood of all
nine BCI IV-2a subjects at five trials per class and reduced the error of the joint transition matrix by 61\%.
Decoding accuracy did not improve ($\bm{p=0.17}$). Cross-subject priors therefore stabilize
few-trial dynamic models of motor-imagery EEG, but they do not improve decoding, which in
these data relies mainly on structure shared across subjects.
\end{abstract} 


\section{Introduction}
\IEEEPARstart{I}{magery} of a hand movement desynchronizes the sensorimotor $\mu$ (8 to
13~Hz) and $\beta$ (13 to 30~Hz) rhythms, most strongly over the hemisphere contralateral
to the imagined hand~\cite{pfurtscheller1999,pfurtscheller2001}. The effect is transient,
varies from trial to trial and unfolds in both hemispheres at once. Hidden Markov models
(HMMs) capture such dynamics with a small number of discrete states and have been used to
classify single motor-imagery trials~\cite{obermaier2001} and to describe large-scale
brain dynamics in magnetoencephalography and functional imaging~\cite{vidaurre2017,vidaurre2018}.
When several signals interact, a single HMM over their joint state space becomes large, and
coupled models are preferable. The influence model~\cite{asavathiratham2001} and its
generalization, the latent structure influence model (LSIM), represent each signal by its
own Markov chain and let the state of every chain depend on the previous states of all
chains through a convex combination of pairwise transition matrices. Karimi and
Shamsollahi derived tractable inference~\cite{karimi2020} and an exact maximum likelihood
(ML) expectation-maximization (EM) algorithm for the LSIM and demonstrated it on EEG and
electrocorticography~\cite{karimi2023}.

The number of LSIM parameters grows quadratically with both the number of chains and the
number of states,
while motor-imagery recordings rarely contain more than a few dozen trials per class and
subject. In the PhysioNet EEG Motor Movement/Imagery Dataset~\cite{schalk2004,goldberger2000}
a subject contributes about 22 trials per imagined hand. Subject-specific ML estimates are
then dominated by sampling noise and by the local optima of EM, and the large
inter-subject variability of sensorimotor rhythms~\cite{saha2020} makes it unclear how much
information can be shared. Transfer learning addresses the same shortage for
discriminative decoders. Regularized and composite common spatial patterns shrink
subject-specific spatial filters toward those of other subjects~\cite{lotte2011,kang2009},
subject-independent classifiers are trained on pooled data~\cite{fazli2009}, and Riemannian
methods align covariance matrices across sessions and subjects~\cite{zanini2018}. Reviews
document many further variants~\cite{jayaram2016,wu2022,lotte2018}. For generative
state-space models two strategies dominate. In speech recognition, MAP adaptation shrinks
the parameters of a speaker-specific HMM toward those of a speaker-independent model through
conjugate priors~\cite{gauvain1994}. In neuroimaging, a group-level HMM is fitted to data pooled
across subjects~\cite{vidaurre2018}, and subject-specific versions can then be re-estimated
from the group solution, which is called dual estimation~\cite{gohil2024}. Neither strategy has been
formulated for the LSIM, whose selector variables and pairwise transition matrices do not
appear in a standard HMM.

This paper develops a MAP estimator for the LSIM with a cross-subject (population) prior
and validates it on two public motor-imagery datasets. The contributions are as follows.
\begin{enumerate}
\item A conjugate prior for all LSIM parameters, defined as the expected complete-data
sufficient statistics of $\lambda$ pseudo-trials from a population model. The resulting MAP
M-step is in closed form, no EM iteration decreases the log posterior, and the single
strength parameter $\lambda$ has a direct interpretation as a number of trials.
\item A leave-one-subject-out empirical Bayes construction~\cite{morris1983} of the
population model, with a canonical state ordering that puts subject and population states in
correspondence.
\item A validation protocol fixed before testing~\cite{nosek2018}. The prior strength was
chosen on 35 tuning subjects of a public EEG dataset, and the estimator was then evaluated
once on 70 held-out subjects of the same dataset and on a second public dataset with a prior
learned from the first. A simulation with known parameters complements these evaluations.
\end{enumerate}

\section{Latent Structure Influence Model of the Two Sensorimotor Rhythms}
\subsection{Coupled Markov Chains for the Left and Right Sensorimotor Cortex}
An LSIM has $C$ chains. Chain $c$ has a hidden state $q_t^c\in\{1,\dots,M\}$ and emits an
observation $\bo_t^c\in\mathbb{R}^L$ at frame $t=1,\dots,T$. Given the joint state
$q_{t-1}=(q_{t-1}^1,\dots,q_{t-1}^C)$, the chains evolve conditionally independently according to
\begin{equation}
P\bigl(q_t^c=j\mid q_{t-1}\bigr)=\sum_{c'=1}^{C}\theta_{c'c}\,a^{c'c}_{q_{t-1}^{c'}\,j},
\label{eq:lsim}
\end{equation}
where $\theta_{c'c}\ge0$ with $\sum_{c'}\theta_{c'c}=1$ is the influence of chain $c'$ on
chain $c$, and $\mathbf{A}^{c'c}=[a^{c'c}_{ij}]$ is a row-stochastic $M\times M$ matrix. The
initial states are independent with $P(q_1^c=n)=\pi_{cn}$. Emissions are Gaussian,
\begin{equation}
p\bigl(\bo_t^c\mid q_t^c=n\bigr)=\Nc\bigl(\bo_t^c\mid\bmu_{cn},\bSig_{cn}\bigr).
\label{eq:emis}
\end{equation}
In this study $C=2$. Chain 1 describes the left sensorimotor cortex (Laplacian derivation
C3$'$) and chain 2 the right (C4$'$). The observation of each chain at frame $t$ is the pair of
log $\mu$ and log $\beta$ powers ($L=2$), and each chain has $M=2$ states: a desynchronized
state with low power and a state with higher power. The weight $\theta_{c'c}$ expresses how
strongly the previous state of hemisphere $c'$ drives the next state of hemisphere $c$.
The sum in \eqref{eq:lsim} has an equivalent latent form. A selector $s_t^c\in\{1,\dots,C\}$
is drawn with $P(s_t^c=c')=\theta_{c'c}$, and chain $c$ then moves according to row
$q_{t-1}^{c'}$ of $\mathbf{A}^{c'c}$. With $K$ independent trials
$\mathcal{O}=\{\bo_{k,1:T}\}_{k=1}^{K}$ and parameters
$\Lambda=\{\bm\theta,\{\mathbf{A}^{c'c}\},\bm\pi,\{\bmu_{cn},\bSig_{cn}\}\}$, the complete-data
log-likelihood is
\begin{multline}
\log p(\mathcal{O},\mathcal{Q},\mathcal{S}\mid\Lambda)=\sum_{k,c}\log\pi_{c\,q_{k,1}^c}
+\sum_{k,c}\sum_{t=2}^{T}\Bigl[\log\theta_{s_{k,t}^c c}\\
+\log a^{s_{k,t}^c c}_{q_{k,t-1}^{s}\,q_{k,t}^c}\Bigr]
+\sum_{k,c,t}\log\Nc\bigl(\bo_{k,t}^c\mid\bmu_{c\,q_{k,t}^c},\bSig_{c\,q_{k,t}^c}\bigr),
\label{eq:cll}
\end{multline}
where $q^{s}_{k,t-1}$ abbreviates the previous state of the selected chain
$s=s_{k,t}^c$. Only the product of $\theta_{c'c}$ and $\mathbf{A}^{c'c}$ enters
\eqref{eq:lsim}, so the two are not separately identifiable~\cite{karimi2023}. The quantities
reported below are therefore either functions of the joint transition matrix
\begin{equation}
P_{\mathbf{i}\mathbf{j}}=\prod_{c=1}^{C}\sum_{c'=1}^{C}\theta_{c'c}\,a^{c'c}_{i_{c'}j_c},
\qquad \mathbf{i},\mathbf{j}\in\{1,\dots,M\}^C,
\label{eq:joint}
\end{equation}
which is identifiable up to a relabeling of states, or the effective influence
$I_{c'\to c}=\theta_{c'c}\max_{i,i'}\dTV(\mathbf{a}^{c'c}_{i\cdot},\mathbf{a}^{c'c}_{i'\cdot})$,
which combines both factors into a single measure.

\subsection{Exact E-Step and Expected Transition Counts}
The E-step runs the scaled forward-backward recursion on the product chain with $M^C$
states and transition matrix \eqref{eq:joint}. From the two-slice posteriors it yields, for
each chain pair and state pair, the expected selector-transition counts
\begin{equation}
\xi^{c'c}_{ij}=\sum_{k}\sum_{t=2}^{T}P\bigl(s_{k,t}^c=c',\,q_{k,t-1}^{c'}=i,\,q_{k,t}^c=j\mid\mathcal{O},\Lambda^{\mathrm{old}}\bigr),
\label{eq:xi}
\end{equation}
together with the state posteriors $\gamma_{k,t}^{cn}=P(q_{k,t}^c=n\mid\mathcal{O},\Lambda^{\mathrm{old}})$,
the occupancies $w_{cn}=\sum_{k,t}\gamma_{k,t}^{cn}$ and the initial counts
$r_{cn}=\sum_k\gamma_{k,1}^{cn}$. The selector posterior in \eqref{eq:xi} follows from the
joint two-slice posterior by multiplying with $\theta_{c'c}a^{c'c}_{ij}$ and dividing by the
transition probability of chain $c$~\cite{karimi2023}.

\section{Population-Informed Maximum a Posteriori Estimation}
\subsection{Prior Defined by Pseudo-Trials From a Population Model}
Let $\bar\Lambda$ be a population model fitted to the trials of other subjects (Section
\ref{sec:pop}). One E-step of $\bar\Lambda$ on its own training data, with the resulting
counts divided by the number of pooled trials, gives the expected statistics of a single
population trial,
$\bar\xi^{c'c}_{ij}$, $\bar w_{cn}$ and $\bar\pi_{cn}$, together with the population emission
parameters $\bar\bmu_{cn}$ and $\bar\bSig_{cn}$. For a strength $\lambda\ge0$ the prior is
\begin{multline}
p(\Lambda)\propto\prod_{c',c}\theta_{c'c}^{\,\lambda\sum_{ij}\bar\xi^{c'c}_{ij}}
\prod_{c',c,i,j}\bigl(a^{c'c}_{ij}\bigr)^{\lambda\bar\xi^{c'c}_{ij}}
\prod_{c,n}\pi_{cn}^{\,\lambda\bar\pi_{cn}}\\
\times\prod_{c,n}|\bSig_{cn}|^{-\kappa_{cn}/2}
\exp\Bigl\{-\tfrac{\kappa_{cn}}{2}\operatorname{tr}\bigl[\bSig_{cn}^{-1}\bigl(\bar\bSig_{cn}
+\mathbf{d}_{cn}\mathbf{d}_{cn}^{\top}\bigr)\bigr]\Bigr\},
\label{eq:prior}
\end{multline}
with $\kappa_{cn}=\lambda\bar w_{cn}$ and $\mathbf{d}_{cn}=\bmu_{cn}-\bar\bmu_{cn}$. The
transition and initial factors are Dirichlet densities with parameters $1+\lambda\bar\xi$ and
$1+\lambda\bar\pi$. The emission factor is a normal-inverse-Wishart density with mean
$\bar\bmu_{cn}$, scale $\kappa_{cn}$, scale matrix $\kappa_{cn}\bar\bSig_{cn}$ and
$\kappa_{cn}-L-2$ degrees of freedom, which is proper when $\kappa_{cn}>2L+1$.

\begin{proposition}[Pseudo-data equivalence]
Up to an additive constant,
$\log p(\Lambda)=\lambda\,\E_{\bar\Lambda}\bigl[\log p(\bo_{1:T},q_{1:T},s_{2:T}\mid\Lambda)\bigr]$,
where the expectation is the per-trial expected complete-data log-likelihood
\eqref{eq:cll} evaluated with the population statistics.
\end{proposition}
\begin{proof}
Equation \eqref{eq:cll} is linear in the counts of selector transitions, initial states and
state occupancies, so its expectation replaces them by $\bar\xi$, $\bar\pi$ and $\bar w$.
For the emissions,
$\E[(\bo-\bmu)^{\top}\bSig^{-1}(\bo-\bmu)]=\operatorname{tr}[\bSig^{-1}(\bar\bSig+\mathbf{d}\mathbf{d}^{\top})]$
when the pseudo-observations of state $n$ have mean $\bar\bmu_{cn}$ and scatter
$\bar\bSig_{cn}$. This holds at a fixed point of the ML updates of the population model
whenever its eigenvalue floor is inactive. Comparing terms with the logarithm of
\eqref{eq:prior} gives the result.
\end{proof}

The MAP estimate therefore maximizes the likelihood of the subject's $K$ trials augmented by
$\lambda$ virtual trials whose statistics are those of the population. With $T=50$ frames,
$\lambda=1$ contributes 50 pseudo-frames per chain, spread over states in proportion to the
population occupancy.

\subsection{Closed-Form M-Step and Convergence of MAP-EM}
The E-step is unchanged. The M-step maximizes the expected complete-data log-likelihood plus
$\log p(\Lambda)$. Each parameter block is conjugate, and the maximizers are
\begin{align}
\theta_{c'c}&=\frac{\sum_{ij}\bigl(\xi^{c'c}_{ij}+\lambda\bar\xi^{c'c}_{ij}\bigr)}
{\sum_{c''}\sum_{ij}\bigl(\xi^{c''c}_{ij}+\lambda\bar\xi^{c''c}_{ij}\bigr)},\label{eq:mtheta}\\
a^{c'c}_{ij}&=\frac{\xi^{c'c}_{ij}+\lambda\bar\xi^{c'c}_{ij}}{\sum_{j'}\bigl(\xi^{c'c}_{ij'}+\lambda\bar\xi^{c'c}_{ij'}\bigr)},\label{eq:ma}\\
\pi_{cn}&=\frac{r_{cn}+\lambda\bar\pi_{cn}}{K+\lambda},\label{eq:mpi}\\
\bmu_{cn}&=\frac{\sum_{k,t}\gamma_{k,t}^{cn}\bo_{k,t}^c+\kappa_{cn}\bar\bmu_{cn}}{w_{cn}+\kappa_{cn}},\label{eq:mmu}\\
\bSig_{cn}&=\frac{\mathbf{S}_{cn}+\kappa_{cn}\mathbf{d}_{cn}\mathbf{d}_{cn}^{\top}+\kappa_{cn}\bar\bSig_{cn}}{w_{cn}+\kappa_{cn}},\label{eq:msig}
\end{align}
where $\mathbf{S}_{cn}=\sum_{k,t}\gamma_{k,t}^{cn}(\bo_{k,t}^c-\bmu_{cn})(\bo_{k,t}^c-\bmu_{cn})^{\top}$
uses the updated mean. The mean \eqref{eq:mmu} does not depend on $\bSig_{cn}$, so the pair
\eqref{eq:mmu} and \eqref{eq:msig} is the joint maximizer. As in~\cite{karimi2023}, the
eigenvalues of $\bSig_{cn}$ are floored at $\epsilon=10^{-3}$. Eigenvalue clipping is the
exact constrained maximizer of a block of the form
$-\frac{a}{2}\log|\bSig|-\frac12\operatorname{tr}(\mathbf{B}\bSig^{-1})$, which is the form of
the emission term with $a=w_{cn}+\kappa_{cn}$.

\begin{proposition}[Ascent and limits]
No MAP-EM iteration decreases $\log p(\mathcal{O}\mid\Lambda)+\log p(\Lambda)$.
For $\lambda=0$ the updates \eqref{eq:mtheta} to \eqref{eq:msig} reduce to the ML updates of
\cite{karimi2023}. For $\lambda\to\infty$ the estimate converges to the population model
$\bar\Lambda$.
\end{proposition}
\begin{proof}
The E-step is exact and the M-step maximizes the expected complete-data log posterior
exactly within the feasible set, so the ascent argument of Dempster \emph{et
al.}~\cite{dempster1977} for MAP estimation applies. The case $\lambda=0$ is immediate.
As $\lambda$ grows, every ratio in \eqref{eq:mtheta} to \eqref{eq:msig} is dominated by the
population terms, which equal the population parameters because $\bar\Lambda$ is a fixed
point of the ML updates on its own data.
\end{proof}

With $K$ trials, the counts of the subject carry a weight proportional to $K$ and those of
the population a weight proportional to $\lambda$. This is the classical shrinkage of
empirical Bayes~\cite{morris1983} and of MAP adaptation~\cite{gauvain1994}.

\subsection{Leave-One-Subject-Out Population Model and State Matching}\label{sec:pop}
For a target subject, the population model of each class was fitted by ML to the pooled
trials of that class from all other contributing subjects. Each contributing subject was
standardized with its own mean and standard deviation per feature, and the target subject
with the statistics of its own training trials, so that the population and the subject share
the same scale. The prior requires state $n$ of the subject to correspond to state $n$ of the
population. All models were initialized by $k$-means with the states of each chain ordered by
the mean of the first feature ($\mu$ power), and after fitting, the states of the population
model were relabeled into the same order. State 1 of every chain is therefore the
low-$\mu$-power, desynchronized state.

\subsection{Control Estimator With an Uninformative Prior}
To separate the effect of population information from that of regularization alone, a flat
prior of the same form was defined with uniform statistics, $\bar\xi^{c'c}_{ij}=(T-1)/(CM^2)$,
$\bar w_{cn}=T/M$, $\bar\pi_{cn}=1/M$, $\bar\bmu_{cn}=\mathbf{0}$ and
$\bar\bSig_{cn}=\mathbf{I}$, which is appropriate for standardized features. The
corresponding estimator is called Flat-MAP, and the estimator with the population prior is
called Group-MAP.

\subsection{Complete Estimation Procedure}
Algorithm~\ref{alg:map} lists the steps for one subject and one class. The cost of an EM
iteration is $\mathcal{O}(KTM^{2C})$ for the forward-backward recursion, as for ML, because
the prior only adds constants to the expected counts.

\begin{algorithm}[t]
\caption{Population-informed MAP estimation of an LSIM for one target subject and one class.}
\label{alg:map}
\begin{algorithmic}[1]
\Require training trials of the target subject, same-class trials of the other subjects, prior strength $\lambda$
\State standardize each other subject with its own feature means and standard deviations
\State fit the population model $\bar\Lambda$ by ML-EM to the pooled trials of the other subjects
\State order the states of every chain of $\bar\Lambda$ by mean $\mu$ power
\State run one E-step of $\bar\Lambda$ on the pooled trials and divide by their number to obtain $\bar\xi$, $\bar w$, $\bar\pi$, $\bar\bmu$ and $\bar\bSig$
\State standardize the target subject with the statistics of its training trials
\State initialize $\Lambda$ by $k$-means with states ordered by mean $\mu$ power
\Repeat
\State E-step: forward-backward recursion on the product chain, giving $\xi$, $\gamma$, $w$ and $r$
\State M-step: update $\bm\theta$, $\mathbf{A}$, $\bm\pi$, $\bmu$ and $\bSig$ by \eqref{eq:mtheta} to \eqref{eq:msig}
\State floor the eigenvalues of every $\bSig_{cn}$ at $10^{-3}$
\Until{the log posterior rises by less than $10^{-6}$ per frame, or 150 iterations}
\Ensure MAP estimate $\hat\Lambda$
\end{algorithmic}
\end{algorithm}

\section{EEG Database and Validation Design}
\subsection{Motor-Imagery Recordings From Two Public Datasets}
The EEG Motor Movement/Imagery Dataset (EEGMMIDB)~\cite{schalk2004,goldberger2000} contains
recordings of 109 volunteers made with the BCI2000 system from 64 electrodes of the
international 10-10 system at a sampling rate of 160~Hz. In runs 4, 8 and 12 a target
appeared on the left or the right side of a screen, and the subject imagined opening and
closing the corresponding fist until the target disappeared. The 327 EDF files of these runs
(version 1.0.0) were retrieved from PhysioNet, and every file matched the checksum published
with the dataset.
Subjects S088, S092 and S100 were excluded because their recordings were sampled at 128~Hz,
and S104 because of irregular trial durations, which leaves 105 subjects. A trial was
rejected when the largest peak-to-peak amplitude of the ten electrodes entering the
Laplacian derivations (1 to 40~Hz) exceeded the subject median by more than three scaled
median absolute deviations. This removed 142 of 4725 trials and left 18 to 24 trials per
class and subject.

The BCI Competition IV data set 2a~\cite{brunner2008,tangermann2012} contains EEG of nine
subjects recorded with 22 electrodes at 250~Hz during cued imagery of the left hand, right
hand, feet and tongue. Left-hand and right-hand trials of the first session without expert
artifact marks were used, which gives 53 to 69 trials per class and subject. The recordings were read
from a public NumPy conversion of the original GDF files, in which every training session
contains the documented 72 trials per class. This dataset differs from
EEGMMIDB in amplifier, montage, sampling rate, cue and subject population, and it served
only as an external test set.

\subsection{Laplacian $\mu$ and  $\beta$ Band-Power Features}
Small Laplacian derivations~\cite{hjorth1975,mcfarland1997} were formed at C3 and C4 with
the neighbors FC3, C5, C1, CP3 and FC4, C2, C6, CP4, which exist in both montages. Each
derivation was band-pass filtered in the $\mu$ (8 to 13~Hz) and $\beta$ (13 to 30~Hz) bands
with fourth-order zero-phase Butterworth filters. The squared magnitude of the analytic signal
was averaged in non-overlapping 100-ms bins and log-transformed. A trial spans 1~s before to
4~s after the cue, which gives $T=50$ frames. The model has $C=2$ chains (C3$'$ and C4$'$),
$M=2$ states per chain and $L=2$ features per chain ($\log\mu$ and $\log\beta$ power). One
model was estimated per class. Features were standardized per subject with the mean and
standard deviation of the training trials.

\subsection{Compared Estimators and Fitting Settings}
Four estimators were compared. ML is the EM algorithm of~\cite{karimi2023} ($\lambda=0$).
Flat-MAP uses the generic prior with strength $\lambda_{\mathrm{flat}}$. Group-MAP uses the
population prior with strength $\lambda_{\mathrm{group}}$. Pooled is the population model
itself and uses no data of the target subject. All subject-level fits started from the same
$k$-means initialization and ran at most 150 iterations, with a stopping tolerance of
$10^{-6}$ per frame on the log posterior.

\subsection{Analysis Plan}
The analysis plan, including all settings and outcomes, was fixed before the test subjects
and BCI IV-2a were analyzed. The 105 EEGMMIDB subjects were randomly divided into a tuning set
of 35 subjects and a test set of 70 subjects. Table~\ref{tab:sources} summarizes the analyses, their data and the numbers of subjects
and training trials.

\begin{table}[t]
\centering
\caption{Summary of the analyses. Trials are training trials per class and subject.}
\label{tab:sources}
\setlength{\tabcolsep}{3pt}
\footnotesize
\begin{tabular}{l l c c}
\toprule
Analysis & Data & Subjects & Trials\\
\midrule
Selection of $\lambda$ & EEGMMIDB & 35 & 14 to 20\\
Primary outcomes & EEGMMIDB & 70 & 15 to 20\\
Reduced training sets & EEGMMIDB & 70 & 5, 10\\
Split-half reliability & EEGMMIDB & 70 & 9 to 12\\
External validation & BCI IV-2a & 9 & 5 to 30\\
Exploratory analysis & EEGMMIDB & 105 & 5, 14 to 20\\
Simulation & simulated & 35 & 5 to 40\\
\bottomrule
\end{tabular}
\end{table}

\subsubsection{Choice of the Prior Strength on 35 Tuning Subjects}
For each tuning subject the population model was built from the other 34 tuning subjects.
Both prior strengths were chosen from the grid $\lambda\in\{0.25,0.5,1,2,4,8,16,32,64\}$
pseudo-trials by maximizing the mean held-out log-likelihood per frame
\begin{equation}
\bar\ell=\frac{1}{TK_{\mathrm{te}}}\sum_{k\in\mathrm{te}}\log p\bigl(\bo_{k,1:T}\mid\hat\Lambda_{y_k}\bigr)
\label{eq:ell}
\end{equation}
computed with 5-fold stratified cross-validation, where $\mathrm{te}$ denotes the held-out
trials, $K_{\mathrm{te}}$ their number and $\hat\Lambda_{y_k}$ the fitted model of the true
class $y_k$ of trial $k$.
The selected values were stored and used unchanged in every later analysis.

\subsubsection{Evaluation on 70 Held-Out Subjects}
For each test subject the population model was built from all other 104 subjects. In 5-fold
stratified cross-validation each estimator was trained on all training trials of a fold (15
to 20 per class) and on random subsets of 5 and 10 trials per class. Three primary outcomes
were defined for all training trials and corrected together by the Holm
procedure~\cite{holm1979}: the difference in $\bar\ell$ between Group-MAP and ML (P1), the
difference in $\bar\ell$ between Group-MAP and Flat-MAP (P2) and the difference in decoding
accuracy between Group-MAP and ML (P3). A trial was decoded as the class whose model gave the
larger likelihood. Secondary outcomes were the same contrasts at 5 and 10 trials per class,
the comparison with Pooled, the decoding accuracy of log-variance features with linear
discriminant analysis (LDA) as an external reference, and the split-half reliability of
subject-level summaries. For
the latter, odd and even trials of each class were fitted separately, and the Spearman
correlation across subjects between the two halves was computed for the stationary
occupancy of the desynchronized state of each chain, the effective influences
$I_{\mathrm{R}\to\mathrm{L}}$ and $I_{\mathrm{L}\to\mathrm{R}}$, the $\mu$-power difference
between the two states of each chain, and the class difference of the asymmetry
$I_{\mathrm{R}\to\mathrm{L}}-I_{\mathrm{L}\to\mathrm{R}}$. Paired contrasts were tested with
the two-sided Wilcoxon signed-rank test and summarized by the mean, a bootstrap 95\%
confidence interval (CI) with 10\,000 resamples and the number of subjects improved.

\subsubsection{External Validation on BCI IV-2a}
For each BCI IV-2a subject and class, a fixed random split separated a pool of 30 trials from
a test set with the remaining 23 to 39 trials. Features were standardized with the pool.
Training sets of 5, 10, 20 and 30 trials per class were drawn from the pool, with 20
independent draws for each size below 30. The prior was built from all 105 EEGMMIDB subjects
and therefore contained no BCI IV-2a data. A secondary prior was built from the other eight BCI IV-2a
subjects. The outcomes were $\bar\ell$ on the test trials, decoding accuracy, and the squared
errors of the joint transition matrix and of the emission means relative to an independent
reference model fitted by ML to the test trials. Because the reference is independent of the
training data, differences in expected squared error between estimators are unaffected by
the noise of the reference.

\subsubsection{Simulation With Known Parameters}
Ground truths were the ML fits of the 35 tuning subjects to their left-hand trials. For each
synthetic subject, 22 trials were simulated from each of the other 34 truths to build its
population model, and training sets of 5, 10, 20 and 40 trials and a test set of 300 trials
were simulated from its own truth. The truths therefore carry the full heterogeneity of the recorded
subjects, including the estimation noise of their own fits, and the prior is consequently
misspecified.
The outcomes were the excess held-out loss
$\frac{1}{TK}\sum_k[\log p(\bo_k\mid\Lambda^{\star})-\log p(\bo_k\mid\hat\Lambda)]$, which
estimates the Kullback-Leibler divergence rate from the true model $\Lambda^\star$, and the
squared Frobenius error of the joint transition matrix \eqref{eq:joint}.

\section{Results}
\subsection{Prior Strength Selected on the Tuning Subjects}
\begin{figure*}[t]
\centering
\includegraphics[width=\textwidth]{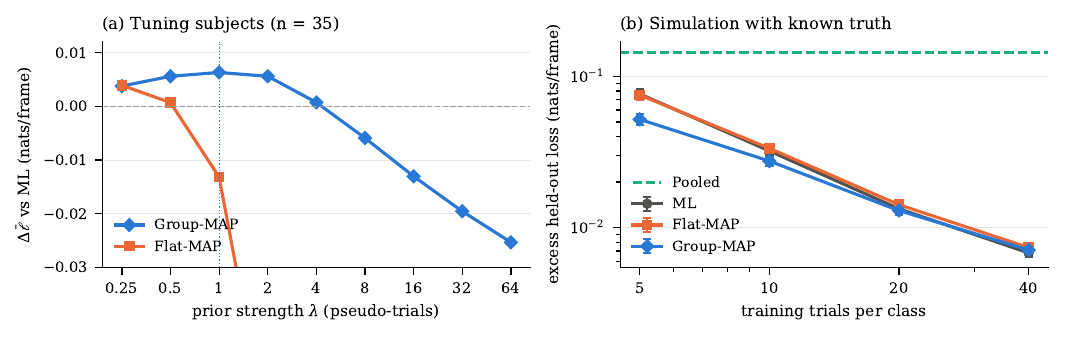}
\caption{(a) Mean change in held-out log-likelihood per frame relative to ML on the 35 tuning
subjects as a function of the prior strength. The dotted line marks the selected
$\lambda=1$. Flat-MAP continues below the axis for $\lambda\ge2$. (b) Excess held-out loss
relative to the true model in the simulation with ground truths fitted to the tuning subjects
(mean and standard error over 35 synthetic subjects, logarithmic axes) for the selected prior
strengths.}
\label{fig:tune}
\end{figure*}

Fig.~\ref{fig:tune}(a) shows the mean held-out log-likelihood of the tuning subjects relative
to ML. Group-MAP improved on ML for $\lambda\le4$ and peaked at $\lambda=1$ with a gain of
0.0063 nats per frame. Stronger priors pulled the subject models too far toward the
population, and at $\lambda=64$ the loss reached 0.025 nats per frame. Flat-MAP gained 0.0039
nats per frame at $\lambda=0.25$ and deteriorated quickly beyond $\lambda=0.5$, falling to
$-0.061$ at $\lambda=2$. The selected values were $\lambda_{\mathrm{group}}=1$ and
$\lambda_{\mathrm{flat}}=0.25$. The pooled model alone was 0.036 nats per frame worse than ML,
so subject-specific information mattered when all training trials were available.

\subsection{Held-Out Likelihood in the 70 EEGMMIDB Test Subjects}
\begin{figure}[t]
\centering
\includegraphics[width=\columnwidth]{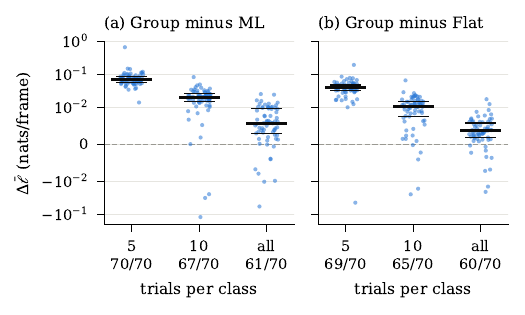}
\caption{Per-subject differences in held-out log-likelihood per frame for the 70 test
subjects with 5, 10 and all (15 to 20) training trials per class. The axis is symmetric
logarithmic and linear within $\pm0.01$. Horizontal bars mark the median and quartiles, and
the counts under each group give the subjects with a positive difference.}
\label{fig:mmidb}
\end{figure}

Table~\ref{tab:mmidb} and Fig.~\ref{fig:mmidb} summarize the test stage. With all training
trials, Group-MAP increased $\bar\ell$ over ML by 0.0051 nats per frame (95\% CI 0.0026 to
0.0072) and improved the likelihood of 61 of 70 subjects. It exceeded Flat-MAP by 0.0031
(95\% CI 0.0018 to 0.0043) in 60 of 70 subjects. After Holm correction the primary outcomes P1 and P2 were confirmed
($p=9.1\times10^{-8}$ and $p=2.5\times10^{-7}$). The gains grew as the training set shrank.
With ten trials per class the improvement over ML was 0.019 nats per frame (67 of 70
subjects), and with five trials it was 0.082 nats per frame and occurred in all 70 subjects.
At five trials Group-MAP also exceeded Flat-MAP in 69 of 70 subjects. Since a trial has 50
frames, 0.082 nats per frame corresponds to 4.1 nats per trial.

The comparison with Pooled depended on the amount of subject data. With all training trials
Group-MAP was better in 51 of 70 subjects. With ten trials each estimator was better in 35
subjects.
With five trials the pooled model gave a higher likelihood in 50 of 70 subjects
($p=8.8\times10^{-4}$), although the mean difference still favored Group-MAP because the
pooled model failed badly for a minority of subjects whose dynamics differ from those of the
population.

\begin{table*}[t]
\centering
\caption{Results for the 70 EEGMMIDB test subjects: held-out log-likelihood per frame $\bar\ell$ (nats), paired differences with bootstrap 95\% CIs, numbers of subjects improved, two-sided Wilcoxon $p$-values (unadjusted) and decoding accuracies.}
\label{tab:mmidb}
\setlength{\tabcolsep}{3.2pt}
\footnotesize
\resizebox{\textwidth}{!}{%
\begin{tabular}{l cccc ccc ccc}
\toprule
Trials & \multicolumn{4}{c}{$\bar\ell$} & \multicolumn{3}{c}{Group-MAP minus ML} & \multicolumn{3}{c}{Group-MAP minus Flat-MAP}\\
\cmidrule(lr){2-5}\cmidrule(lr){6-8}\cmidrule(lr){9-11}
per class & ML & Flat-MAP & Group-MAP & Pooled & mean [95\% CI] & better & $p$ & mean [95\% CI] & better & $p$\\
\midrule
5 & $-$5.4837 & $-$5.4443 & $-$5.4021 & $-$5.4121 & 0.0816 [0.0689, 0.1019] & 70/70 & $3.6\times10^{-13}$ & 0.0422 [0.0366, 0.0487] & 69/70 & $2.1\times10^{-12}$\\
10 & $-$5.3847 & $-$5.3772 & $-$5.3657 & $-$5.4121 & 0.0190 [0.0133, 0.0237] & 67/70 & $4.0\times10^{-10}$ & 0.0115 [0.0091, 0.0141] & 65/70 & $9.5\times10^{-11}$\\
15 to 20 & $-$5.3523 & $-$5.3503 & $-$5.3472 & $-$5.4121 & 0.0051 [0.0026, 0.0072] & 61/70 & $3.0\times10^{-8}$ & 0.0031 [0.0018, 0.0043] & 60/70 & $1.2\times10^{-7}$\\
\midrule
 & \multicolumn{5}{c}{accuracy (\%)} & \multicolumn{3}{c}{Group-MAP minus ML (points)} & & \\
\cmidrule(lr){2-6}\cmidrule(lr){7-9}
 & ML & Flat-MAP & Group-MAP & Pooled & LDA & mean [95\% CI] & better/worse & $p$ & & \\
\midrule
5 & 55.7 & 55.8 & 55.8 & 60.6 & 58.2 & 0.1 [$-$1.2, 1.4] & 33/26 & 0.98\\
10 & 56.8 & 56.2 & 57.1 & 60.6 & 57.6 & 0.2 [$-$0.9, 1.4] & 31/23 & 0.61\\
15 to 20 & 56.2 & 56.6 & 57.2 & 60.6 & 60.3 & 1.0 [$-$0.3, 2.3] & 30/21 & 0.17\\
\bottomrule
\end{tabular}}
\end{table*}

\subsection{Left-Hand Versus Right-Hand Decoding Accuracy}
The third primary outcome was not confirmed. With all training trials Group-MAP decoded
57.2\% of the test trials correctly, and ML decoded 56.2\% (difference 1.0 points, 95\% CI $-0.3$ to
2.3, Holm-adjusted $p=0.17$). Group-MAP did not differ from ML with 10 or 5 trials
per class either ($p=0.61$ and $p=0.98$). Two reference decoders performed better. Log-variance features with LDA
reached 60.3\%. The pooled class models, which never saw the test subject, reached 60.6\% and
exceeded Group-MAP in 44 of 70 subjects ($p=0.012$). With five trials per class the advantage
of the pooled models over Group-MAP grew to 4.9 points ($p=3.0\times10^{-4}$).

\subsection{Split-Half Reliability of Subject-Level Model Summaries}
Table~\ref{tab:rel} lists the split-half reliabilities. The prior did not detectably
change the reliability of any summary, and all confidence intervals of the difference
included zero. The largest change was an increase from 0.33 to 0.45 for the occupancy of the
desynchronized state at C4$'$. The class difference of the hemispheric asymmetry remained
unreliable with both estimators (0.26 and 0.19).

\begin{table}[t]
\centering
\caption{Split-half reliability (Spearman $\rho$ across the 70 test subjects) of subject-level summaries, and bootstrap 95\% CI of the difference Group-MAP minus ML.}
\label{tab:rel}
\setlength{\tabcolsep}{3pt}
\footnotesize
\begin{tabular}{l c c c}
\toprule
Summary & ML & Group-MAP & difference\\
\midrule
desynchronized-state occupancy, C3$'$ & 0.56 & 0.58 & [$-$0.06, 0.10]\\
desynchronized-state occupancy, C4$'$ & 0.33 & 0.45 & [$-$0.04, 0.28]\\
$I_{\mathrm{R}\to\mathrm{L}}$ & 0.17 & 0.22 & [$-$0.18, 0.27]\\
$I_{\mathrm{L}\to\mathrm{R}}$ & 0.31 & 0.33 & [$-$0.21, 0.25]\\
state $\mu$-power difference, C3$'$ & 0.60 & 0.57 & [$-$0.20, 0.16]\\
state $\mu$-power difference, C4$'$ & 0.68 & 0.58 & [$-$0.28, 0.07]\\
class difference of asymmetry & 0.26 & 0.19 & [$-$0.29, 0.14]\\
\bottomrule
\end{tabular}
\end{table}

\subsection{Transfer of the EEGMMIDB Prior to BCI IV-2a}
\begin{figure*}[t]
\centering
\includegraphics[width=\textwidth]{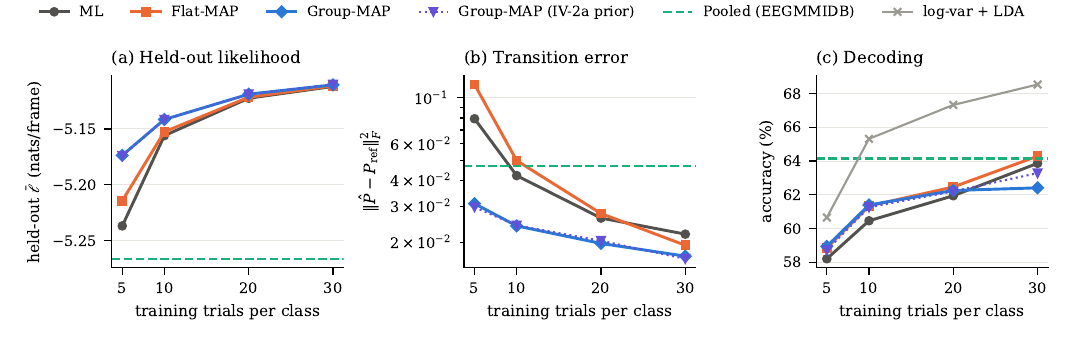}
\caption{External validation on BCI IV-2a as a function of the number of training trials
per class (means over nine subjects). (a) Held-out log-likelihood per frame. (b) Squared
error of the joint transition matrix relative to an independent reference fit (logarithmic
axis). (c) Decoding accuracy. The dashed line is the pooled EEGMMIDB model, which uses no
BCI IV-2a data.}
\label{fig:iv}
\end{figure*}

Table~\ref{tab:iv} and Fig.~\ref{fig:iv} show the results for the second dataset. A prior
learned only from EEGMMIDB improved the held-out likelihood of all nine subjects at five
trials per class ($p=0.004$, the smallest attainable value for nine subjects) and of seven
of nine at ten trials ($p=0.02$). The improvement was smaller and not significant at 20 and
30 trials. The error of the joint transition matrix relative to the independent reference
decreased from 0.079 to 0.031 at five trials (61\%), from 0.042 to 0.024 at ten trials and
from 0.026 to 0.020 at 20 trials, and at each of these sizes the decrease occurred in all
nine subjects. The error of the emission means
decreased from 0.81 to 0.45 at five trials (eight of nine subjects, $p=0.027$). The prior
from the other BCI IV-2a subjects gave results nearly identical to those of the
cross-dataset prior at every size. Decoding accuracy did not differ between Group-MAP and ML at any size. Without any
BCI IV-2a training data, the EEGMMIDB pooled models decoded the BCI IV-2a trials with 64.2\%
accuracy. This exceeded both subject-level estimators at every training size, although the
differences from ML were not significant with nine subjects. LDA on log-variance features was
the best decoder from ten trials onward.

\begin{table*}[t]
\centering
\caption{External validation on BCI IV-2a ($n=9$). Group-MAP uses a prior learned only from EEGMMIDB, and Group-MAP (IV) uses a prior from the other eight BCI IV-2a subjects. The Pooled EEGMMIDB model, which uses no BCI IV-2a data, reached $\bar\ell=-5.2660$, a transition error of 0.0469 and an accuracy of 64.2\%. Values are means over subjects and, for fewer than 30 trials, over 20 training draws.}
\label{tab:iv}
\setlength{\tabcolsep}{3.4pt}
\footnotesize
\resizebox{\textwidth}{!}{%
\begin{tabular}{l cccc cc cc cc ccc}
\toprule
Trials & \multicolumn{4}{c}{held-out $\bar\ell$} & \multicolumn{2}{c}{Group vs ML} & \multicolumn{2}{c}{$\|\hat P-P_{\mathrm{ref}}\|_F^2$} & \multicolumn{2}{c}{Group vs ML} & \multicolumn{3}{c}{accuracy (\%)}\\
\cmidrule(lr){2-5}\cmidrule(lr){6-7}\cmidrule(lr){8-9}\cmidrule(lr){10-11}\cmidrule(lr){12-14}
per class & ML & Flat-MAP & Group-MAP & Group-MAP (IV) & better & $p$ & ML & Group-MAP & better & $p$ & ML & Group-MAP & LDA\\
\midrule
5 & $-$5.2367 & $-$5.2142 & $-$5.1736 & $-$5.1741 & 9/9 & 0.004 & 0.0793 & 0.0308 & 9/9 & 0.004 & 58.2 & 58.9 & 60.7\\
10 & $-$5.1560 & $-$5.1525 & $-$5.1416 & $-$5.1417 & 7/9 & 0.02 & 0.0421 & 0.0241 & 9/9 & 0.004 & 60.5 & 61.4 & 65.3\\
20 & $-$5.1226 & $-$5.1217 & $-$5.1190 & $-$5.1190 & 8/9 & 0.10 & 0.0263 & 0.0198 & 9/9 & 0.004 & 61.9 & 62.3 & 67.3\\
30 & $-$5.1121 & $-$5.1115 & $-$5.1106 & $-$5.1103 & 6/9 & 0.25 & 0.0220 & 0.0172 & 6/9 & 0.36 & 63.9 & 62.4 & 68.5\\
\bottomrule
\end{tabular}}
\end{table*}

\subsection{Parameter Recovery in Simulation}
In the simulation (Fig.~\ref{fig:tune}(b)) Group-MAP reduced the excess loss from 0.0765 to
0.0520 nats per frame at five trials, a reduction of 32\% (Wilcoxon $p=3.8\times10^{-5}$), and
from 0.0322 to 0.0276 at ten trials, a reduction of 14\% ($p=0.0034$). At 20 and 40 trials
the two estimators did not differ (0.0133 versus 0.0130, $p=0.44$, and 0.0068 versus 0.0071,
$p=0.73$). The error of the joint transition matrix fell from 0.120 to 0.038 at five trials
and from 0.032 to 0.023 at ten trials, and it was slightly larger for Group-MAP at 40 trials
(0.0069 versus 0.0057). Flat-MAP was close to ML at every size. The pooled model, which
ignores the subject, had an excess loss of 0.144, larger than that of every subject-level
estimator.
The prior strength that minimized the excess loss was 0.5 for 5, 10 and 20 trials and 0.25
for 40 trials, close to the value selected on the tuning subjects.

\subsection{Exploratory Analysis of Likelihood and Decoding Across Prior Strengths}
\begin{figure}[t]
\centering
\includegraphics[width=\columnwidth]{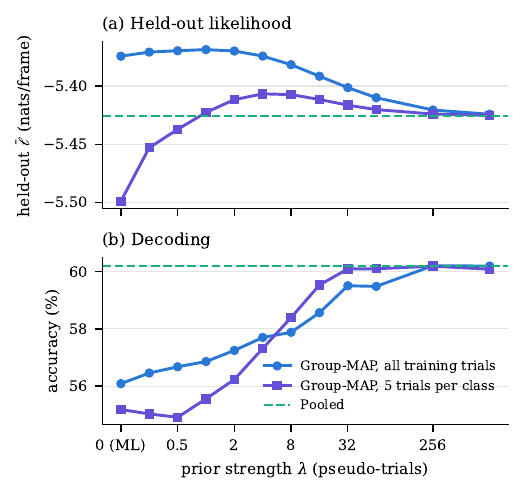}
\caption{Exploratory analysis on all 105 EEGMMIDB subjects. (a) Held-out log-likelihood per frame and (b) decoding accuracy of Group-MAP as functions of the prior strength (ML isplotted at the left end). The dashed line marks the pooled models, which use no data of the
target subject.}
\label{fig:explore}
\end{figure}

The following analysis was not part of the analysis plan. It was added after the test stage
had shown that the pooled models decoded better than every subject-level estimator, and its
results should be read as exploratory. Group-MAP was refitted over the whole grid of prior
strengths, extended to $\lambda=256$ and $\lambda=1024$, for all 105 EEGMMIDB subjects with
population models from all other subjects (Fig.~\ref{fig:explore}). Held-out likelihood and
decoding accuracy peaked at different strengths. With all training trials $\bar\ell$ was
highest at $\lambda=1$, whereas accuracy rose monotonically from 56.1\% for ML to 60.2\% for
$\lambda\ge256$, where it equaled the accuracy of the pooled models (60.2\%). With five trials per class the
likelihood optimum moved to $\lambda=4$, and accuracy again approached the pooled value only
for very strong priors. No prior strength gave better decoding than the pooled models.

\section{Discussion}
\subsection{Effect of the Population Prior on Few-Trial Models}
The population prior made subject-level LSIMs predict unseen trials better, and the gain
became smaller as the amount of subject data increased. With five trials per class the improvement
occurred in every EEGMMIDB test subject and in every BCI IV-2a subject, and the error of the
joint transition matrix fell by more than half in both the simulation and the external
dataset. With 15 to 20 trials the gain was small but still consistent across subjects, and
with 30 to 40 trials it vanished. This is the pattern expected from shrinkage, which
reduces variance at the cost of bias and pays off mainly when the subject contributes few
trials. The simulation, in which the truths were models fitted to
recorded subjects and therefore more heterogeneous than the prior assumes, showed a small cost of the
fixed $\lambda$ for the transition matrix at 40 trials. A prior strength chosen once on
tuning subjects is thus a reasonable default for few-trial data but not for long recordings.

Two controls locate the source of the gain. Flat-MAP, which applies the same conjugate
machinery with uninformative statistics, improved little over ML and degraded quickly as its
strength increased. Group-MAP exceeded it in 60 of 70 subjects with all trials and in 69 of
70 with five. The benefit therefore comes from the information in the population model, not
from regularization alone. The pooled model, at the other extreme, was worse than Group-MAP
for 51 of 70 subjects when all training trials were used, so the subject data still carried
information that the population lacked.

The cross-dataset result is of practical interest. A prior learned from 105 EEGMMIDB subjects
helped BCI IV-2a subjects as much as a prior learned from the other BCI IV-2a subjects,
although the two datasets differ in amplifier, montage, sampling rate, cue and subject
population.
After per-subject standardization, the state dynamics of Laplacian $\mu$ and $\beta$ power at
C3 and C4 appear to be similar enough across these settings for a population model to be
reused. Studies with few subjects could therefore borrow a population model from a large
open dataset.

\subsection{Why the Prior Did Not Improve Decoding}
The prior strength was selected for predictive likelihood, and the likelihood of a 5-s trial
is dominated by temporal structure shared by both classes, such as the dwell times of the
desynchronized states and the coupling between the hemispheres. Shrinking both class models
toward their population counterparts by one pseudo-trial improves this shared structure but
hardly changes the difference between the class models, which is what the likelihood-ratio
decoder uses. The pooled class models, in contrast, decoded better than every
subject-specific estimate in EEGMMIDB and transferred to BCI IV-2a without any calibration.
The class-discriminative part of these features, essentially the contralateral
desynchronization, is thus largely common to subjects and is estimated poorly from 15 to 20
trials of a single subject. This agrees with the success of subject-independent decoders for
sensorimotor rhythms~\cite{fazli2009} and with the transfer literature~\cite{jayaram2016,wu2022}.
For decoding, a population-informed generative model would need either a much stronger prior
or discriminative training. A simple log-variance decoder remained competitive or better once
ten or more trials per class were available. The exploratory analysis supports this reading. Accuracy increased steadily as the subject
models were pulled toward the population, while the likelihood had its optimum at a weak
prior, so a prior strength chosen for one goal is not optimal for the other.

\subsection{Implications for Interpreting Subject-Level Model Summaries}
A better predictive model did not produce more reliable subject-level summaries. The
split-half reliabilities of occupancies, influences and state contrasts were unchanged within
their confidence intervals, and the class difference of the hemispheric asymmetry stayed
below 0.3 with both estimators. Shrinkage moves all subjects toward the same center, which
leaves their rank order, and hence a rank correlation, largely intact. Between-subject
comparisons of LSIM coupling measures estimated from about 20 trials per class should
therefore be interpreted with caution, whichever of the two estimators is used.

\subsection{Limitations and Extensions}
A single strength $\lambda$ was shared by all parameter blocks and all subjects. The tuning
curve and the simulation suggest that the optimal strength depends on the number of trials
and on how typical a subject is. Hierarchical estimation of separate strengths for the
dynamics and the emissions, or of subject-specific strengths by marginal likelihood, is a
natural extension. Each fit used a single $k$-means initialization, so part of the remaining
variance may come from local optima of EM. The model was deliberately small, with two chains,
two states and two features per chain at C3 and C4, and Gaussian emissions. The feature
settings and the model size were fixed in a preceding analysis of the BCI IV-2a data and were
not tuned here, so BCI IV-2a is independent of the prior and its strength but not of these
settings. The external
dataset has only nine subjects, which limits the power of its tests, and all analyses were
offline. The analysis plan was fixed in advance but was not publicly registered.

\section{Conclusion}
A conjugate prior built from the expected statistics of a population model turns the exact
EM algorithm for the latent structure influence model into a MAP estimator with closed-form
updates, monotone ascent and a single interpretable strength parameter. Under a protocol
fixed in advance, the estimator improved the held-out likelihood of motor-imagery EEG in 61
of 70 held-out subjects with all training trials and in all 70 with five trials per class.
It also reduced parameter error in a simulation with known parameters, and a prior learned
from one dataset improved the models of another. The prior did not improve decoding or the reliability of
subject-level summaries, and population class models alone decoded better than
subject-specific ones. Population information is therefore most useful for stabilizing
few-trial descriptions of sensorimotor dynamics, while the class-discriminative part of the
signal is already largely shared across subjects.

\section*{Data and Code Availability}
EEGMMIDB is publicly available through PhysioNet under the Open Data Commons Attribution License v1.0, while the BCI Competition IV Dataset 2a is available through the official competition website. The source code, analysis plan, and result files have been publicly released.

\end{document}